\documentclass[book,numbers]{elsbook-P04}

\usepackage{amsmath,amssymb,amsfonts,amsthm,makeidx,graphicx,colortbl,chapterbib,natbib,algorithm,subfigure}

\def\mSigma{\mbox{$\mathbf{\Sigma} \kern .08em$}}
\def\mLambda{\mbox{$\mathbf{\Lambda} \kern .08em$}}

\def\bfs{\mbox{\boldmath $f_s$}}
\newcommand{\ds}{\displaystyle}

\def\bL{\text{\mbox{\boldmath $L$}}}

\def\b0{\text{\mbox{\boldmath $0$}}}
\def\ba{\text{\mbox{\boldmath $a$}}}

\def\bp{\text{\mbox{\boldmath $p$}}}

\def\bu{\text{\mbox{\boldmath $u$}}}

\begin{document}

\begin{frontmatter}%%% Chapter Opener

\chapter[Chapter Title]{The edge cloud: A holistic view of communication, computation and caching}\label{chap1}

\author*[1]{Sergio Barbarossa}%
\author[1]{Stefania Sardellitti}%
\author[1]{Elena Ceci}%
\author[1]{Mattia Merluzzi}%

%\address[2]{\orgname{}, \orgdiv{Department of Engineering}, \orgaddress{Via G. Duranti 93, 06125, Perugia, Italy.}}
\address[1]{\orgname{Sapienza University of Rome}, \orgdiv{Dept. of Information Engineering, Electronics, and Telecommunications}, \orgaddress{via Eudossiana 18, 00184, Rome, Italy.}}
\address*[3]{Corresponding: \email{sergio.barbarossa@uniroma1.it}}

\makechaptertitle

\begin{abstract}[Abstract]
The evolution of communication networks shows a clear shift of focus from just improving the communications aspects to enabling new important services, from Industry 4.0 to automated driving, virtual/augmented reality, Internet of Things (IoT), and so on. This trend is evident in the roadmap planned for the deployment of the fifth generation (5G) communication networks. This ambitious goal requires a paradigm shift towards a vision that looks at communication, computation and caching $(3C)$ resources as three components of a single holistic system. The further step is to bring these $3C$ resources closer to the mobile user, at the {\it edge} of the network, to enable very low latency and high reliability services.  The scope of this chapter is to show that signal processing techniques can play a key role in this new vision.  In particular, we motivate the joint optimization of $3C$ resources. Then we show how graph-based representations can play a key role in building effective learning methods and devising innovative resource allocation techniques.
\end{abstract}

\begin{keywords}[Keywords:] 5G networks, wireless communications, graph-based learning
\end{keywords}

\end{frontmatter}%

\section{Introduction}
The major goal of next generation (5G) communication networks is to build a communication infrastructure that will enable new business opportunities in diverse sectors, or {\it verticals}, such as automated driving, e-health, virtual/augmented reality, Internet of Things (IoT), smart grids, and so on  \cite{whitepaper2016}, \cite{andrews2014will}. These services have very different specifications and requirements in terms of latency, reliability, data rate, number of connected devices, and so on. Thinking of enabling such diverse services using a common communication platform might then look like a crazy idea. But, in reality, if the system is properly designed, reusing a common infrastructure for different purposes might induce a significant economic advantage. The key idea for making this possible  is to use {\it virtualization} \cite{mijumbi2016network} and implement {\it network slicing} \cite{rost2017network}. Through virtualization, many network functionalities are implemented in software through virtual machines that can be instantiated and moved upon request \cite{vassilaras2017algorithmic}. Building on virtualization, network slicing partitions a {\it physical} network into multiple {\it virtual} networks, each matched to its specific requirements and constraints, thus enabling operators to provide networks on an as-a-service basis, while meeting a wide range of use cases in parallel.

This new reality, sometimes called fourth industrial revolution, can be realized by a new architecture able to meet advanced requirements, especially in terms of latency (below $5$ ms), reliability (around 0.99999), coverage (up to $100\,$ devices$/m^2$), and data rate (more then 10 Gbps).  At the physical layer, 5G builds on a significant increase of system capacity by incorporating massive MIMO techniques, dense deployment of radio access points, and wider bandwidth. All these strategies are facilitated by the introduction
of millimeter wave (mmWave) communications \cite{heath2016overview}, \cite{xiao2017millimeter}, \cite{sakaguchi2017}: mmWaves make possible the reduction of the antenna size, thus enabling the use of array with many elements, as required in massive MIMO; dense deployment is also facilitated because mmWaves give rise to a stronger intercell attenuation; finally, increasing the carrier frequency facilitates the usage of wider bandwidths. However, the significant improvement achievable at the physical layer could be still insufficient to meet the challenging and diverse requirements of very low latency and ultra reliability. A further improvement comes from a paradigm shift that puts applications at the center of the system design. Network Function Virtualization (NFV) and Multi-access Edge Computing (MEC) \cite{taleb2017multi} are the key tools of this application-centric networking. In particular, MEC plays the key role of bringing cloud-computing resources at the edge of the network, within the Radio Access Network (RAN), in close proximity to mobile subscribers \cite{taleb2017multi}, \cite{ETSI_white_paper_2015}. MEC is particularly effective to deliver context-aware services or to enable computation offloading from resource-poor mobile devices to fixed servers or to perform intelligent cache pre-fetching, based on local learning of the most popular contents across space and time.\\

Given this perspective, the goal of this chapter is to show that graph-based methods can play a significant role in optimizing resource allocation or deriving new learning mechanisms. The organization of this chapter is the following.  In Section \ref{Holistic view of communication, computation and caching} we present the edge-cloud architecture and we motivate the holistic approach that looks at $3C$ resources as a common pool of resources to be handled jointly with the goal of achieving, on the user side,  a satisfactory quality of experience and, on the network side, a balanced and efficient use of resources.  Then, in Section \ref{Joint optimization of communication and computation}, we will focus on the joint optimization of computation and communication resources, with specific attention to computation offloading in the edge-cloud. In Section \ref{Joint optimization of caching and communication}, we will concentrate on the joint optimization of caching ad communication.  Differently from storage, which is fundamentally {\it static}, caching is inherently {\it dynamic}, so that cache memories are pre-fetched when and where needed, and then released. In both cases of joint optimization, the goal is to bring resources, either computation (virtual machines) or cache, as close as possible to the end user, to enable truly low latency and low energy consumption services.
After presenting this holistic view, we will move in Section \ref{Graph-based learning} to present some learning mechanisms based on graph signal processing. In particular, we show how to reconstruct the radio environment map (REM), which enables a cognitive usage of the radio resources. % In Section \ref{Graph-based resource allocation} we will show how to manage network resource allocation based again on graph representations.
Then, building again on graph representations, in Section \ref{Network reliability}, we show how to achieve an optimal resource allocation across a network while being robust to link failures. The proposed approach is based on a small perturbation analysis of network topologies affected by sporadic edge failures. Finally, in Section \ref{Conclusions} we draw some conclusions and suggest some possible further developments.

\section{Holistic view of communication, computation and caching}
\label{Holistic view of communication, computation and caching}
\begin{figure}[h]
\centering
\includegraphics[width=10cm]{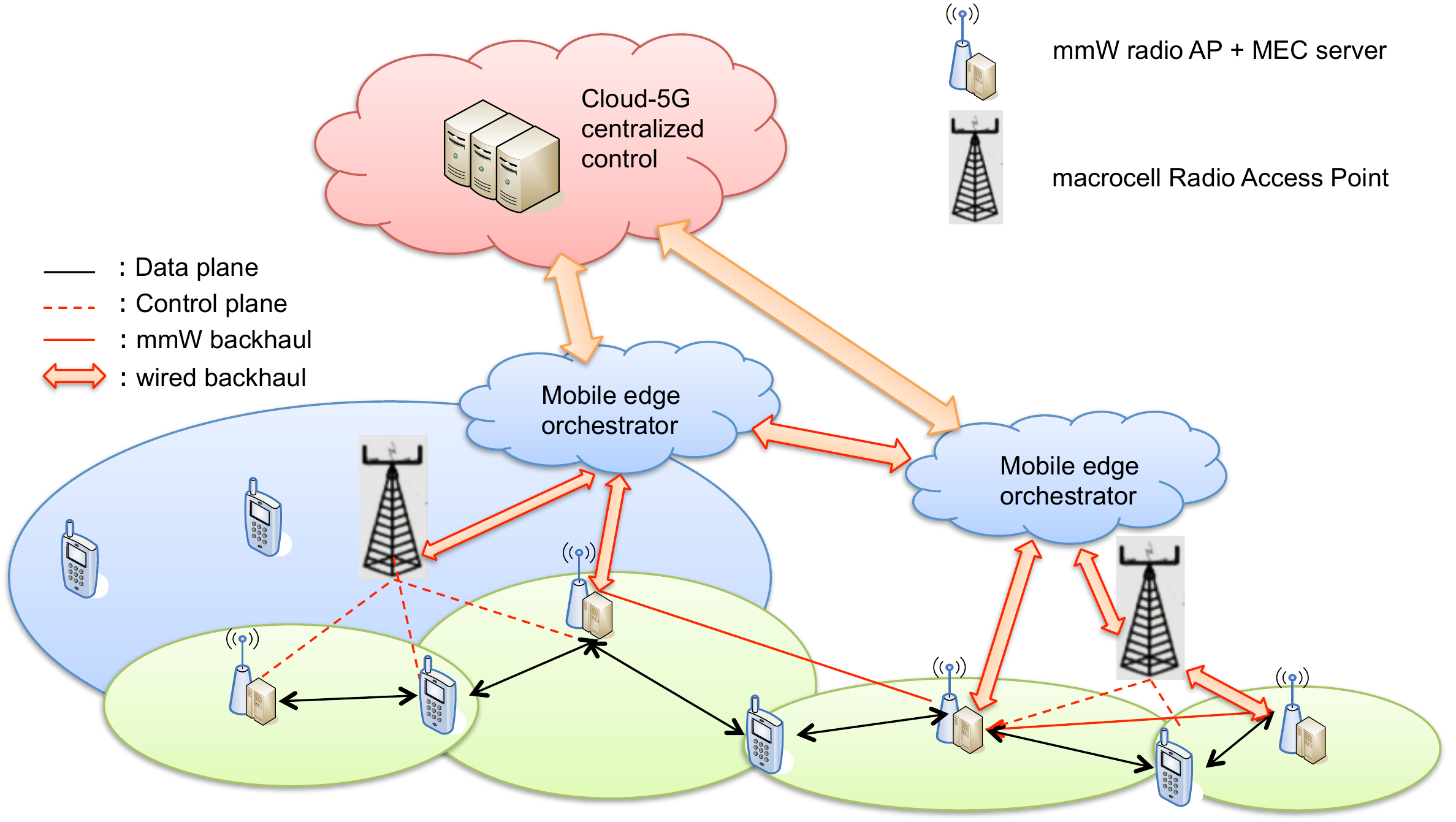}
\caption{Edge-cloud architecture.}
\label{fig:edge-cloud}
\end{figure}

The new infrastructure provided by next communication networks can be seen as a truly distributed and pervasive computer that provides very different services to mobile users with sufficiently good quality of experience. The physical resources composing this pervasive computer are cache memories, computing machines, and communication channels. The system should serve the end user, either a mobile subscriber or a car or the component of a production process with, ideally, zero latency, which means an end-to-end latency smaller than the user perception capability or than the maximum value ensuring proper control, like breaking time in automated driving. To enable this vision, at the physical layer, the network will support a much higher system (or area) capacity (bits/sec/km$^2$). In 5G systems, a $1,000$-fold increase of system capacity is planned, exploiting mmWave communications, massive MIMO, and dense deployment of access points. However, in spite of this enormous improvement in system capacity, the zero-latency ideal could still be far to be obtained because it is very complicated, if not impossible, to control latency over a wide area network. For this reason, the next step is to  bring computation and cache resources as close as possible to the end user, where proximity is actually measured in terms of service time. This creates a new eco-system, called edge-cloud, whose architecture is sketched in Fig. \ref{fig:edge-cloud}. In this system, within a macro-cell served by one base station, we have multiple millimeter-wave access points (AP), covering much smaller areas.
Each AP is endowed with computation and caching capabilities, to enable mobile users to get proximity access to cloud functionalities. This makes possible to provide  cloud services with very low latency and high data rate, while at the same time keeping data traffic and computation as local as possible. Of course, the computing and caching capabilities of local MEC servers are significantly lower than a typical cloud, but they also serve a limited number of requests and, whenever their resources are insufficient, they may interact with nearby MEC servers, under the supervision of a MEC orchestrator.
In this system, mobile applications are handled by {\it virtual machines} (or containers) instantiated at the edge of the network, close to the end user. The edge is either the ensemble of network access points, as in Multi-access Edge Computing (MEC) \cite{wang2017survey}, or it might even include the mobile terminals as well, as in {\it fog computing} \cite{bonomi2014fog}.

Similarly, contents move dynamically when and where it is more convenient to have them. Caching can in fact be seen as a {\it non-causal} communication, where content move before they are actually requested, to minimize the downloading time.
In this framework, it makes sense to allocate $3C$ resources {\it jointly}, with the objective of guaranteeing some ultimate user quality of experience.

Assuming such a holistic view perspective, the first important question is why using a {\it common} platform, call it 5G or the generations to come next, to accommodate services having so different requirements, like IoT, virtual reality or automated driving. This is indeed one of the main challenges faced by 5G systems. The  approach proposed in the 5G roadmap is {\it network slicing} \cite{vassilaras2017algorithmic}. A network slice is a virtual network that is implemented on top of a physical network in a way that creates the illusion to
the slice tenant of operating its own dedicated
physical network.

Optimizing network slicing is a first important application of graph-based representations, at a high level. In fact, a mathematical formulation of network slicing has been recently proposed in \cite{zhang2017network}, where the communication network is represented as a graph ${\cal G}=(\cal{V}, \cal{E})$, where $\cal{V}$ is the set of nodes and $\cal{E}$ is the set of directed links. There is a subset of function nodes, enabled with NFV functionalities, that can provide a service function $f$. In general, there are $K$ flows, each requesting a distinct service. The requirement of each service $k$ is represented as a service function chain ${\cal F}(k)$ consisting of a set of functions that have to be performed in the predefined order throughout the network. Zhang et al. in \cite{zhang2017network} formulated the slicing problem as the optimal allocation of service functions across the NFV-enabled nodes, while minimizing the total flow in the network. The problem is a mixed binary linear program, which is NP-hard. Nevertheless, the authors of \cite{zhang2017network} proved that the problem can be relaxed with performance guarantees. This is indeed a very interesting application of a graph-theoretic formulation of a very high-level problem.

In the following two sections, we will focus on the joint optimization of pairs of $3C$ resources, namely
communication and computation in Section \ref{Joint optimization of communication and computation} and communication and caching in Section \ref{Joint optimization of caching and communication}.

\section{Joint optimization of communication and computation}
\label{Joint optimization of communication and computation}
Smartphones have really exploded in their usage and capabilities, placing significant demand upon battery usage. Unfortunately, advancements in battery technology have not kept pace with the demands of users and their smartphones. One approach to overcome the battery energy limitations is to offload computations from mobile devices to fixed devices. Computation offloading may be convenient for the following reasons \cite{barbarossa2014communicating}, \cite{wang2017computation}: i) to save energy and then prolong the battery lifetime of hand-held devices; ii) to enable simple devices, like inexpensive sensors, to run sophisticated applications; iii) to reduce latency.
From a user perspective, one of the parameters mostly affecting the quality of experience is the end-to-end (E2E) latency, i.e. the time necessary to get the result of running an application. In case of offloading, this latency includes: i) the time to send bits from the mobile device to the fixed server to enable the program; ii) the time to run the application remotely; iii) the time to get the result back. It is precisely this E2E latency that couples communication and computation resources and then motivates the joint allocation of these resources. We recall now the approach proposed in \cite{barbarossa2013joint} and later expanded in
\cite{barbarossa2014communicating} and \cite{sardellitti2015joint}.

%Clearly, the convenience of computation offloading is strictly tied to the possibility to provide a proximity radio access to mobile devices. This need matches well with the planned dense deployment of access points.
We first consider  the case where multiple users are served by a single AP/MEC pair. Then, we will move to the more challenging case where multiple users are served by multiple AP's and MEC servers. In the first case, the assignment of each UE to a pair of AP and MEC is supposed to be given; in the second case, the assignment is part of the optimization problem. In both cases, for economical reasons associated to promoting their capillary deployment, the computational capabilities of MEC servers are enormously smaller than a typical cloud. This implies that the number of cores per server is very limited or, in other words, that the available cores in a MEC server must operate in a multi-tasking mode to accommodate the requests of multiple users. This means that a server running $K$ applications for as many mobile users will allocate a certain percentage $\beta_k$ of its CPU time to the users that are being served concurrently. If $F_S$ denotes the number of CPU cycles/sec that the server can run, the percentage of CPU cycles/sec assigned to the $k$-th user is then $f_k=\beta_k F_S$.\\

\noindent
{\it Multiple users served by a single AP/MEC pair}\\

\noindent
We start by considering $K$ user equipments (UE) assigned to a single AP and a single MEC. The decision to offload a computation from the mobile device to the MEC server depends on the characteristics of the application to be offloaded. Not all applications are  equally amenable to offloading. The decision should take into account all sources of energy consumption in a smartphone, like display, network, CPU, GPS, camera, and so on. Profiling energy consumption of applications running on smartphones, rather than on a general purpose computer, is not an easy task because of   {\it asynchronous power behavior}, where the effect on a component's power state due to a program
entity lasts beyond the end of that program entity
\cite{pathak2012energy}. The signal processing community could provide a significant contribution to this research field by optimizing app developments taking into account the associated energy profiling for a class of smartphone operating systems, e.g. OS, Android, and so on, and a class of applications. In this chapter, we do not dig into these aspects. We rather concentrate on the joint optimization of radio and computational resources associated to computation offloading, in a multiuser context. From this point of view, we simplify the classification of applications by identifying a few most significant parameters, as relevant for computation offloading. For each user $k$, we consider: i) the number $b_{k}$ of  bits to be transmitted from the mobile user to the server to transfer the program execution; ii) the number of CPU cycles $w_{k}$ necessary to run the application to be offloaded. We denote by $L_{k}$ the E2E latency requested from UE $k$.
The overall latency $T_{k}$ experienced by the $k$-th UE for offloading an application is the sum of three terms: i) the time ${T}^{\texttt{tx}}_{k}$ necessary to transmit all bits to the server to enable the transfer of program execution; ii) the time $T^{\texttt{exe}}_{k}$ for the server to run the application; iii) the time $T^{\texttt{rx}}_{k}$ to get the result back to the UE. In formulas,
\begin{equation}\label{Avg_Delay_single_server}
T_{k}={T}^{\texttt{tx}}_{k}+T^{\texttt{exe}}_{k}+T^{\texttt{rx}}_{k}.
\end{equation}
This equation, in its simplicity, shows that enforcing an E2E latency constraint induces a coupling between communication and computation resources.\\

From a user-centric perspective, the goal might either be to minimize the E2E latency, under a maximum transmit power constraint or, by duality, to minimize the transmit power necessary to guarantee a desired latency. We follow this latter approach, but clearly the two strategies can be interchanged. Let us now express the single contributions in (\ref{Avg_Delay_single_server}) in terms of the parameters to be optimized.

The first contribution is the time ${T}^{\texttt{tx}}_{k}$ to transmit $b_k$ bits from the UE to the AP:
\begin{equation}
T^{\texttt{tx}}_{k}(p_k)=\displaystyle \frac{c_{k}}{r_{k} (p_k)}
\end{equation}
where $c_{k}=b_k/B$, $B$ is the bandwidth and $r_{k}(p_k)$ is the spectral efficiency over the channel between UE and AP, which is equal to
\begin{equation} \label{rate}
r_{k}(p_k)= \log_2\
\left(1+\alpha_{k} p_k \right)
\end{equation}
where ${p}_k$ is the transmit power of UE $k$; $\alpha_{k}=|{h}_{k}|^2/(d_{k}^{\gamma} \sigma_n^2)$ is a an equivalent channel coefficient
that incorporates the channel coefficient ${h}_{k}$, the noise variance $\sigma_n^2$, the distance $d_k$ between UE and AP, and the channel exponent factor $\gamma$. The second contribution in (\ref{Avg_Delay_single_server}) is the execution time at the server, which is equal to $T^{\texttt{exe}}_{k}=w_k/f_k$.  From the user perspective, the third term in (\ref{Avg_Delay_single_server}) does not imply a transmit power, but only the energy to process the received data. This term is typically much smaller than the first term and in the following derivations we will assume it to be a fixed term incorporated in the overall latency.

We are now ready to formulate the computation offloading optimization problem in terms of the transmit powers $p_k$ and the CPU percentages $f_k$, $k=1, \ldots, K$:
%The overall (average) latency associated to computation offloading from the $k$-th MUE is
%\begin{equation}
%\label{Avg_Delay}
%L'_{k}={\Delta}_{tk}+\frac{w_k}{f_{k}}+T_{rk}
%\end{equation}
%where ${w_k}/f_{k}$ is the time necessary to run $w_k$ instructions at the server side and
%$T_{rk}$ denotes the time necessary for the server to send the results back to the
%MUE.  This last time does not depend on parameters associated to the MUE and we
%incorporate it in the latency limit by introducing, for the sake of simplicity of notation,
%the variable $L_k:=L'_{k}-T_{rk}$. Denoting by $\bp:=(p_1, \ldots p_K)$
%and $\bfs:=(f_{1}, \ldots f_K)$ the transmit powers and the computation percentages
%associated to the MUE's, the optimization problem can then be cast as follows:
\begin{equation}
\begin{array}{llll}
\underset{\bp,  \bfs}{\min}
&  \ds \sum_{k=1}^{K} p_{k},  \quad \quad \quad \quad \quad \quad [\mathbf{P.1}]\\
\vspace{0.3cm}
\mbox{s.t.} & \ds \frac{c_k}{ \log_2\left(1+ p_k \alpha_k\right)}+\ds \frac{w_k}{f_{k}}\leq {L}_{k}, \;k=1,\ldots,K\\
& 0 < p_k \leq P_{T},\quad f_{k}>0,  \quad \; k=1,\ldots,K\\
  & \ds \sum_{k=1}^{K}f_{k}\le F_{S}\\
\end{array} \;
\label{average_rate}
\end{equation}
where $\mathbf{p}=(p_1, \ldots, p_K)$ and  $\mathbf{f_s}=(f_1, \ldots, f_K)$.\\

This is a convex problem that can be easily solved. In particular, the optimal computational rates can be expressed in closed form as \cite{barbarossa2017overbooking}:
\begin{equation}
f_k=\frac{\sqrt{w_k\,\eta_k}}{\sum_{k=1}^K\,\sqrt{w_k\eta_k}}F_S,
\end{equation}
where $\eta_k$ are coefficients that depend on the channel coefficients. This simple formula shows how the allocation of computational resources depends not only on computational aspects, but also on the channel state. Note also that the above formula contrasts with the proportional allocation of computational rates that would have been performed in a conventional system, i.e.
\begin{equation}
f_k=\frac{w_k}{\sum_{k=1}^K{w_k}}\,F_S.
\end{equation}

A further substantial improvement to computation offloading comes from the introduction of mmWave links. Merging MEC with an underlying mmWave physical layer creates indeed a unique opportunity to bring IT services at the mobile user with very low latency and very high data rate. This merge is indeed one of the main objectives of the joint Europe/Japan H2020 Project called 5G-MiEdge (Millimeter-wave Edge Cloud as an Enabler for 5G Ecosystem) \cite{5G-MiEdge}. The challenge coming from the use of mmWave links is that they are more prone to blocking events \cite{andrews2017modeling}, which may jeopardize the benefits of computation offloading. A possible way to counteract blocking events in a MEC system using mmWave links was proposed in \cite{barbarossa2017enabling}, \cite{barbarossa2017overbooking}.\\

\noindent
{\it Multiple users served by multiple AP's and multiple MEC servers}\\

%\subsection{Optimal association of mobile users to access points}
Let us consider now a more complex scenario, where multiple users may get radio access through multiple AP's and multiple MEC's. Besides resource allocation, our goal now is to find also the optimal association between UE's, AP's and MEC servers. We consider a system composed of $N_b$ small cell access points, $N_c$ MEC servers, and  $K$  mobile UE's.  Within the edge-cloud scenario depicted in Fig. \ref{fig:edge-cloud}, the association of a mobile user to an access point does not necessarily follow the same principles of current systems, where a mobile user gets access to the base station with the largest signal-to-noise ratio. In the edge-cloud scenario depicted in Fig. \ref{fig:edge-cloud}, the association of a UE to a pair of AP and MEC server depends not only on radio channel parameters, but also on the availability of computational resources at the MEC server. Furthermore, a UE can get radio access from a certain AP, but its application can run elsewhere, not necessarily on the nearest MEC, depending on the availability of computational resources. Actually, since the applications run as virtual machines (VM), we can think of migrating these VM's in order to follow the user. The orchestration of MEC servers in order to provide seamless service continuity to mobile users is an item that has been recently included in the standardization activities of ETSI, within the MEC study group \cite{ETSI_MEC-018_2017}. Migrating VM's is not an easy task, because the instantiation of a VM requires times that are too large with respect to some of the latency requirements foreseen in 5G. This has motivated significant research efforts in investigating light forms of virtual machines, named {\it containers}, that do not need the instantiation of the whole operating system, but only of a restricted kernel \cite{li2015comparing}.

Here, we do not consider the migration of VM's, but we do consider the possibility of letting a UE get access under one AP, while having its application run in an MEC located elsewhere. In this case, we need to incorporate in the E2E latency the delay along the backhaul link connecting AP and MEC. In particular, we denote by $T_{B n m}$ the latency between  access point $n$ and MEC server  $m$. \\

Following an approach similar to what we proposed in \cite{Sard_Globecom14}, we generalize now the resource allocation problem by incorporating binary variables  $a_{ k n m }\in \{0,1\}$ that assume a value $a_{ k n m }=1$ if user $k$ gets radio access through AP $n$ to have its application running on MEC server $m$, and $a_{ k n m }=0$ otherwise. For the sake of simplicity, we assume that each user is served by a single base station and a single cloud.   Our goal now is to find the optimal assignment rule, together with the optimal transmit powers $p_k$ and the computational rates $f_{mk}$ assigned by MEC server $m$ to UE $k$. As in the previous section, our goal is to minimize the overall UE power consumption, under a latency constraint.\\

%The objective function to be minimized is the sum of the powers spent by each mobile user:
%$$f(\mathbf{p},\mathbf{a}) \triangleq \sum_{k=1}^{K} \sum_{n=1}^{N_b} \sum_{m=1}^{N_c}  p_k  a_{kn m}$$
%where $\mathbf{a}\triangleq (\mathbf{a}_{k})_{k \in \mathcal I}$, $\mathbf{a}_{k}=(a_{kn m})_{\forall n, m}$.
%It can be noted from the latency expression in (\ref{Avg_Delay})
%the interplay between radio access and computational aspects and this calls for a \emph{joint} optimization of the radio resources, the transmit power $\mathbf{p}\triangleq (\mathbf{p}_{k})_{k\in \mathcal I}$ of the MUs and the computational rates $\mathbf{f}\triangleq ({f}_{m k})_{\forall m,k }$.
The resulting  optimization problem  is:
 \begin{equation}
\begin{array}{llll}
\underset{\mathbf{p},  \mathbf{f},\mathbf{a}}{\min}
&   f(\mathbf{p},\mathbf{a}) \triangleq \displaystyle \sum_{k=1}^{K} \sum_{n=1}^{N_b} \sum_{m=1}^{N_c} p_k {a}_{knm}  \quad \quad \quad \quad  (\mathcal{P}) \\  \text{s.t.}
& {\rm i)}\,\, g_{kn m}(p_k,f_{m k}, a_{kn m}) \leq {L}_{k} ,\forall \, k, n,m  \medskip\\
& {\rm ii)}\,\, p_k\leq P_{k},\quad {p}_k\geq 0,\, \forall \, k  \\
& {\rm iii)}\,\, h_{m}(\mathbf{f},\mathbf{a})\triangleq \displaystyle \sum_{k=1}^{K} \sum_{n=1}^{N_b} a_{kn m} f_{mk}\le F_{m}, \; \forall \; m, \; \mathbf{f}\geq \mathbf{0} \\
& {\rm iv)}\,\,  \displaystyle \sum_{n=1}^{N_b} \sum_{m=1}^{N_c} a_{kn m }=1,\; a_{kn m}\in\{0,1\}, \quad \forall \, k,n,m
\end{array} \;
\end{equation}
where $\mathbf{f}:= (f_{mk})_{\forall m,k}$, $\mathbf{a}:= (a_{knm})_{\forall k,n,m}$,  and $$g_{kn m}(p_k,f_{m k}, a_{kn m})\triangleq a_{kn m}\left( \displaystyle \frac{ c_{k}}{r_{kn}(p_k)}+\displaystyle\frac{w_{k}}{f_{ m k}}+ T_{B n m}\right)$$
with $r_{kn}(p_k)= \log_2\
\left(1+\alpha_{kn} p_k \right)$ denoting the spectral efficiency of UE $k$ accessing AP $n$ and $\alpha_{kn}$ the equivalent channel coefficient between UE $k$ and AP $n$.\\

The objective function is the total transmit power consumption from the mobile users. The constraints have the following meaning:
i) the overall latency for each user $k$ must be less than the maximum  value $L_k$;
ii) the total power spent by each user must be lower than a fixed total power budget $P_k$;
iii) the sum of the computational rates $f_{m k}$ assigned by each server cannot exceed the server computational capability  $F_{m}$;
iv) each mobile user should be served by one AP/MEC pair; this is enforced by imposing $ \displaystyle \sum_{n=1}^{N_b} \sum_{m=1}^{N_c}a_{kn m}=1$, for each $k$, together with $a_{kn m}\in\{0,1\}$.\\
% For simplicity we have incorporated the term $T^{rx}_{kn m}$ in the latency limit $L_k$.\\

Unfortunately, problem $\mathcal{P}$ is  a mixed-binary problem and is, in general, NP-hard.
To overcome this difficulty, as we suggested in \cite{Sardellitti2018}, we relax the binary variables $a_{knm}$ to be real variables in the interval $[0, 1]$ and adopt a suboptimal  successive convex approximation strategy \cite{Scutari_PartI}, \cite{Sard_Globecom14}, able to converge to local optimal solutions. Additionally, to drive the assignment variables $a_{knm}$ to contain only one value equal to one and all others to zero, for each $k$, we incorporate a further constraint recently suggested in \cite{zhang2017network}. The penalty method in \cite{zhang2017network}  is based on the fact that the following problem
\begin{equation}
\begin{array}{llll} \label{eq:fact}
\underset{\mathbf{a}_k}{\min}
&   \parallel \mathbf{a}_k +\epsilon \mathbf{1}\parallel_p^{p}\triangleq \displaystyle \sum_{n=1}^{N_b} \sum_{m=1}^{N_c}  (a_{knm}+\epsilon)^p \quad \quad \quad \quad  \\  \text{s.t.}
& \,\, \parallel \mathbf{a}_k\parallel_{1}=1,  \medskip\\
& \,\, a_{k n m} \in [0,1], \; \; \;\forall \,n,m
\end{array} \;
\end{equation}
with $\mathbf{a}_k = (a_{knm})_{\forall n,m}$ and $p \in (0,1)$, $\epsilon>0$, admits an optimal solution that is binary, i.e. only one element is one and all the others are zero.
 The optimal solution is $c_{\epsilon,k}=(1+\epsilon)^p+(N_b N_c-1)\epsilon^p$.
Therefore, by relaxing the binary variables $a_{k n m}$ so that they belong to the following
convex set
\begin{equation} \nonumber
\mathcal A=\{ (\mathbf{a}_k)_{k \in \mathcal I} \; : \; a_{kn m}\in [0,1],\displaystyle \sum_{n=1}^{N_b}  \sum_{m=1}^{N_c} a_{kn m}=1, \forall \, k,n,m \},\vspace{-0.2cm}
\end{equation}
where $\mathcal{I}$ denotes the set of $K$ users, we formulate the following relaxed optimization problem \cite{Sardellitti2018}:
 \begin{equation}
\begin{array}{llll}\label{eq:relaxed}
\underset{\mathbf{p},  \mathbf{f},\mathbf{a}}{\min}
&    f_{P_{\sigma}}(\mathbf{p},\mathbf{a})\triangleq f(\mathbf{p},\mathbf{a}) + \sigma P_{\epsilon}(\mathbf{a})  \quad \quad \quad \quad \quad \quad \quad  (\mathcal{P}_{\sigma}) \\  \text{s.t.}
& {\rm i)}\,\, g_{kn m}(p_k,f_{m k}, a_{kn m}) \leq {L}_{k} ,\forall \, k, n,m  \medskip\\
& {\rm ii)}\,\, h_{m}(\mathbf{f},\mathbf{a})\triangleq \displaystyle \sum_{k=1}^{K} \sum_{n=1}^{N_b} a_{kn m} f_{mk}\le F_{m}, \; \forall \; m, \; \mathbf{f}\geq \mathbf{0} \\
& {\rm iii)} \,\, p_k\leq P_{k},\quad {p}_k\geq 0, \forall \, k \in \mathcal{I}, \,\, \mathbf{a} \in \mathcal{A}  \end{array} \;
\end{equation}
where $\sigma>0$ is the penalty parameter, and
\begin{equation}
P_{\epsilon}(\mathbf{a}) \triangleq \displaystyle \displaystyle \sum_{k=1}^{K}\parallel \mathbf{a}_k +\epsilon \mathbf{1}\parallel_p^{p}-c_{\epsilon,k}.
\end{equation}
It is important to emphasize that this penalty is differentiable with respect to the unknown variables.
Even by relaxing the binary variables $\mathbf{a}$,  problem in (\ref{eq:relaxed}) is still non-convex, since the objective function and the constraints i), ii) are non convex.
In \cite{Sardellitti2018}, we proposed a Successive Convex Approximation (SCA) technique, inspired by \cite{Scutari_PartI}, to devise  an efficient  iterative penalty SCA approximation algorithm (PSCA) converging to a local optimal solution of (\ref{eq:relaxed}). We omit the details here, but we report some numerical results.

\begin{figure}[t]
\centering
\includegraphics[width=7cm]{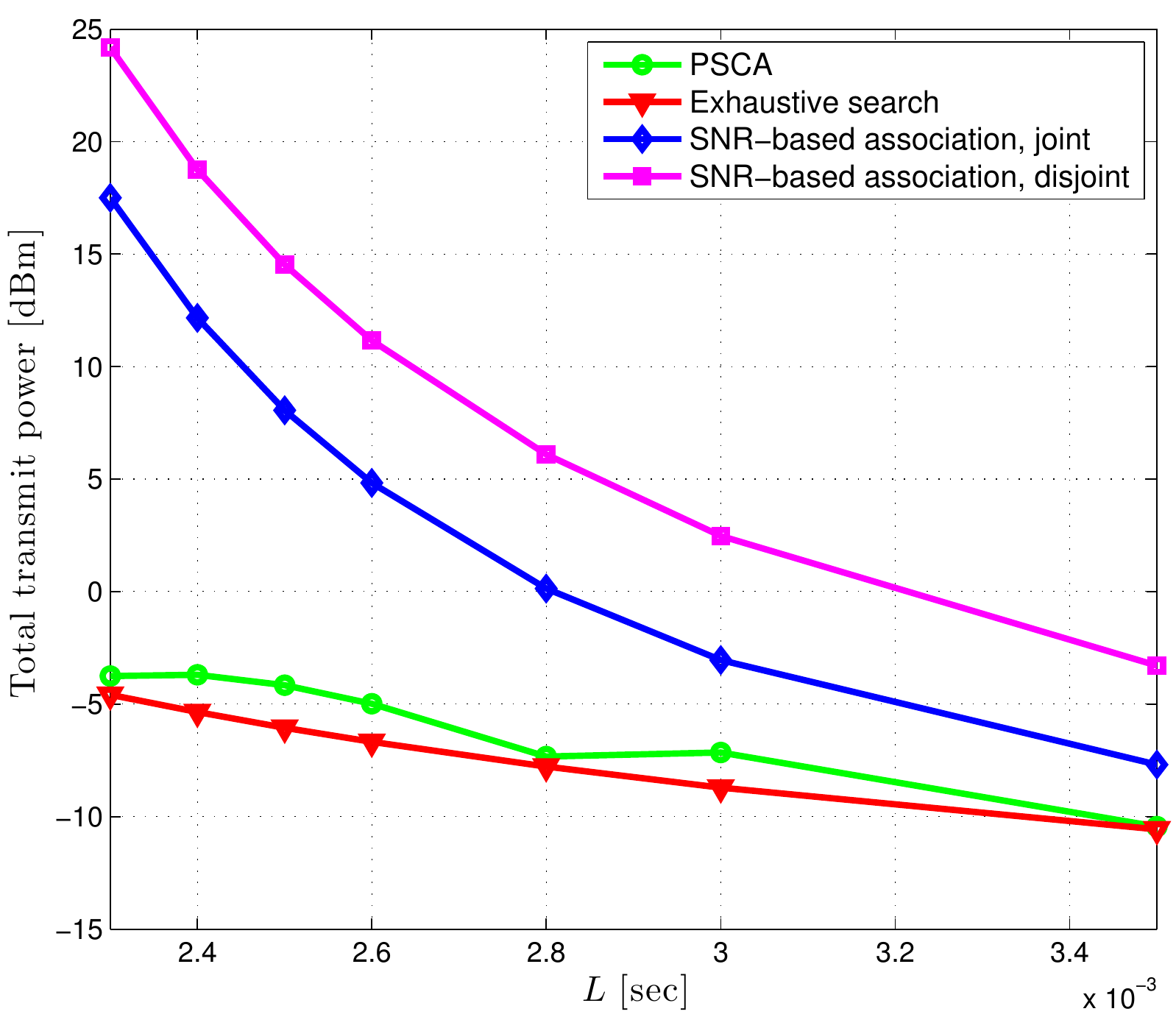}
\caption{Overall UE transmit power consumption vs. $L$.}\label{fig:glob_power}
\end{figure}

To test the effectiveness of the proposed offloading strategy, in Fig. \ref{fig:glob_power} we report
the optimal total transmit power consumption vs. the maximum latency $L_k$. We consider a network composed of $K=4$ users, a number of base stations equal to the
number of clouds, i.e. $N_b=N_c=2$. The other parameters are set as follows: $F_1=2.7\cdot 10^{9}$, $F_2=6 \cdot 10^8$, $P_k=2\cdot 10^{-1}$, $p=0.025$.
From Fig. \ref{fig:glob_power}, we may observe that the PSCA algorithm provides results very close to the exhaustive search algorithm whose complexity is exponential. Additionally, we consider as a comparison term the  SNR-based association method, in both cases where the radio and computational resources are optimized  jointly or disjointly.
It can be noted that the PSCA algorithm  yields considerable power savings  compared to methods based on SNR only, since it  takes advantage of the optimal assignment of
each user to a cloud through the most convenient base station.

\section{Joint optimization of caching and communication}
\label{Joint optimization of caching and communication}
Caching popular contents in storage disks distributed across the network yields significant advantages in terms of reduction of downloading times and limitation of data traffic. Caching can be seen as a {\it non-causal} communication, where popular contents move throughout the network in the off-peak hours to anticipate the users' requests. Clearly an effective caching strategy builds significantly on the ability to learn and predict users' behaviors. This capability lies at the foundation of {\it proactive caching} \cite{bacstuug2015big} and it motivates the need to merge future networks with big data analytics \cite{zeydan2016big}. An alternative approach to proactive caching based on reinforcement learning to learn file popularity across time and space was recently proposed in \cite{sadeghi2017optimal}.\\

Another important pillar of future networks is Information-Centric Networking (ICN), a relatively novel paradigm concerning the distribution of contents throughout the network in a manner much more efficient than conventional Internet  \cite{jacobson2009networking}. Different from what happens in the Internet, where contents are retrieved through their address, in ICN, information is retrieved by {\it named} contents \cite{jacobson2009networking}. In the ICN framework, network entities are equipped with storage capabilities and contents move throughout the network to serve the end user in the best possible way \cite{llorca2013dynamic}. The content placement problem, incorporating number of content copies and their locations in order to minimize a cost function capturing access costs (delay, bandwidth) and/or storage costs, has been formulated as a mixed integer linear program (MILP), shown to be NP-Hard \cite{krishnan2000cache}. In the case where global knowledge of user requests and network resources is available, an Integer Linear Programming (ILP) formulation was given in \cite{llorca2013dynamic}, yielding
the maximum efficiency gains. In this section we recall and extend the formulation of \cite{llorca2013dynamic} to incorporate the cost of inefficient storage of non-popular contents.
\begin{figure}[h]
\centering
\includegraphics[width=6.5cm]{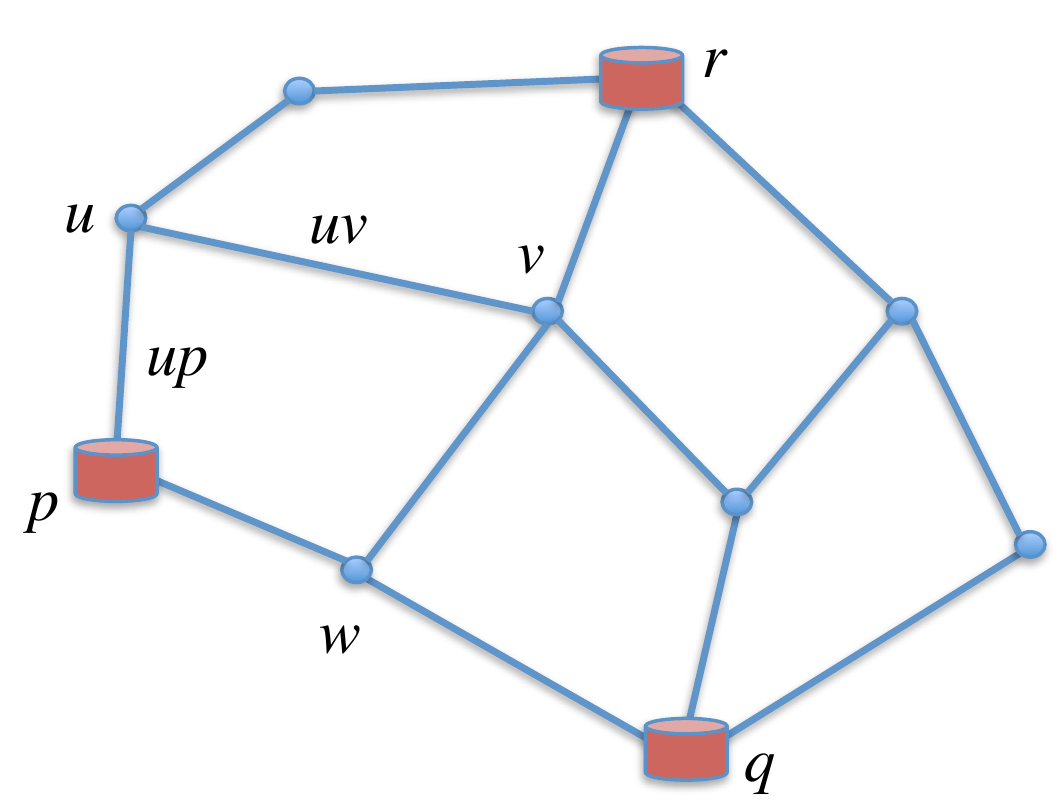}
\caption{Information network.}
\label{fig:ICN}
\end{figure}
Consider an information network ${\cal G}=(\cal{V}, \cal{E}, \cal{K})$, composed of a set of nodes $\cal{V}$, a set of links ${\cal E}$, and a set of information objects ${\cal K}$, as depicted in Fig. \ref{fig:ICN}. A content file can be stored (permanently or temporarily) over the nodes of this graph or travel through its edges. Some contents reside permanently over some repository nodes (e.g., the disks in Fig. \ref{fig:ICN}). In all other nodes (e.g., the circles in Fig. \ref{fig:ICN}), contents may appear and disappear, according to users' requests and network resource allocation. We suppose, for simplicity, that all contents are subdivided into objects of equal size. Each object is then identified by an index $k \in {\cal K}$. Each node is characterized by a storage capability and every edge is characterized by a transport capacity. Time is considered slotted and every slot has a fixed duration $\Delta \tau$. At time slot $n$, each node $u \in {\cal V}$ hosts, as a repository, a set of information objects $K_u[n]\in {\cal K}$ and requests, as a consumer, a set of information objects $Q_u[n]\in {\cal K}$. Let $\mathbf{q}[n] \in \{0,1\}^{|\mathcal{V}||\mathcal{K}|}$ be the request arrival process such that $q_{u}[k,n]=1$
if node $u$ requests object $k$ at time $n$, and  $q_{u}[k,n]=0$  otherwise.

Given this graph, we define a vertex signal over its nodes and an edge signal over its edges.
The vertex signal $s_u[k, n]$ is a binary signal defined as:
\begin{equation}
s_u[k, n]=
\begin{cases}
& 1, \quad {\rm if\,\, content}\,  k, \,{\rm at\,\, time}\, n,\, {\rm is\,\, stored\,\, on\,\,node}\,\, u\\ \nonumber
& 0, \quad \quad \quad \quad \quad {\rm otherwise}\\
\end{cases}, u\in{\cal V}.
\end{equation} The amount of content stored on node $u$, at time $n$, is then $S_u[n]:=\sum_k s_u[k, n]$.
Similarly, we can define an edge signal as a binary signal, defined on each edge, as
\begin{equation}
t_{uv}[k, n]=
\begin{cases}
& 1, \quad {\rm if\,\, content}\,  k, \,{\rm at\,\, time}\, n,\, {\rm is\,\, transported\,\, over\,\,link}\,\, uv\\ \nonumber
& 0, \quad \quad \quad \quad \quad {\rm otherwise}\\
\end{cases}, uv\in{\cal E}.
\end{equation}
The amount of content transported over link $uv$ at time $n$, is then $T_{uv}[n]:=\sum_k t_{uv}[k, n]$. Typically, each content may be host on every node and moved whenever useful. The storage and capacity constraints limit the variability of both $S_u[n]$ and $T_{uv}[n]$ as
\begin{equation}
0 \le S_u[n] \le S_u, \,\,\, 0 \le T_{uv}[n] \le T_{uv},
\end{equation}
where $S_u$ is the storage capability of node $u$, whereas $T_{uv}$ is the transport capacity of link $uv$. The state of the network, at time slot $n$, is represented by the vector $\textbf{x}[n]:=[\textbf{s}[n]; \textbf{t}[n]]$, with $\mathbf{s}[n]:=(s_u[k,n])_{\forall u,k}$ and $\mathbf{t}[n]:=(t_{uv}[k,n])_{\forall k,uv \in \mathcal{E}}$.

In principle, a content $k\in {\cal K}$ may be cached, at any time slot $n$, in more then one location. However, there is a cost in keeping a content in one place, if is not utilized. The goal of \textit{dynamic caching} is to find the state vector $\textbf{x}[n]$ that minimizes an overall cost function that includes the cost for caching and the cost for transportation, under constraints dictated by the storage capability, the transport capacity, and the users' requirements in terms of latency to get access to their desired contents.

The fundamental difference between {\it caching} and {\it storage} is that storage is intrinsically {\it static}, whereas caching is fundamentally {\it dynamic}. This means that cached contents move throughout the network, appear in some nodes and disappear from others. There are only some {\it repository} nodes (e.g., nodes $p, q,$ and $r$ in Fig. \ref{fig:ICN})  that keep a permanent record or have fast access to a content delivery network. The assumption is that each content is host in at least one repository node.

The basic question about caching is then to decide, dynamically, depending on the users' requests, when and where to place all contents, how to move them, and when to drop contents to save memory. The decision for caching an object $k$ at node $u$, at time slot $n$, must result from a trade-off between the cost for storing for a certain amount of time and the cost for transporting the content from its current location to the network access point nearest to the user who requested it.

The cost associated to storing a content $k$ on node $u$ during $T$ consecutive time slots, in the time window $[n'-T+1,n']$, is
\begin{equation}
E_{st}=\sum_{n=n'-T+1}^{n'}\sum_{k\in {\cal K}}\sum_{u\in {\cal V}}s_u[k, n] c_u[k],
\end{equation}
where $c_u[k]$ is the energy cost for keeping content $k$ on node $u$ per unit of time. This unit time cost depends on the popularity of content $k$ in a neighborhood of node $u$. For instance, we can set
\begin{equation}
c_u[k]=\frac{c_0}{1+P_u[k]/P_0}
\end{equation}
where $P_u[k]$ is the popularity of content $k$ at node $u$ and $c_0$ is the (energy) cost for keeping a content object with zero popularity and $P_0$ is the popularity level that justifies halving the cost for caching per unit of time, with respect to zero-popularity contents. The introduction of the cost coefficients $c_u[k]$ is what makes the formulation \textit{context-aware}. In fact, the popularity $P_u[k]$ may vary  across the network.

The cost associated to content transportation is
\begin{equation}
E_{tr}=\sum_{n=n'-T+1}^{n'}\sum_{k\in {\cal K}}\sum_{uv\in {\cal E}}\,t_{uv}[k, n] c_{uv}[k],
\end{equation}
where $c_{uv}[k]$ is the energy cost for transporting object $k$ over link $uv$. In general, when user $u$ makes a request of content $k$, we may associate to that request a maximum delivery time, which we call $D_{u}[k]$. We also denote by ${\cal N}_u$ the neighborhood of node $u$, i.e., the set of nodes that are one hop away from node $u$, and by $\mathbf{x}_T:=[\mathbf{x}[n'-T+1];\ldots;\mathbf{x}[n'] ]$ the state vector during $T$ consecutive time slots.

The dynamic caching optimization problem can then be formulated as
\begin{equation} \label{eq:caching_prob}
\hat{\mathbf{x}}_T={\rm arg}\min_{\mathbf{x}_T} (E_{st}(\mathbf{x}_T)+E_{tr}(\mathbf{x}_T))
\end{equation}
subject to the following constraints
\begin{align}
&(a)\hspace{1cm} q_{u}[k, n] \le s_u[k, n]+\sum_{v\in {\cal N}_u}\sum_{j=0}^{D_{u}[k]}\,t_{vu}[k, n+j]\nonumber\\
&(b) \hspace{1cm} s_u[k, n]  \le  s_u[k, n-1]+\sum_{v\in {\cal N}_u}t_{vu}[k, n-1]\nonumber\\
&(c)  \hspace{1cm} t_{vu}[k, n] \le s_v[k, n-1]+\sum_{w\in {\cal N}_v} t_{wv}[k, n-1]\nonumber\\
&(d)  \hspace{1cm} s_{u}[k, n]  =  1, \forall k \in {K}_u[n],\,\,s_{u}[k, 0]=0, k\notin {K}_u[n]\nonumber\\
&(e)  \hspace{1cm} S_{u}[n]  \le  S_u\nonumber\\
&(f)  \hspace{1cm} T_{uv}[n]   \le  T_{uv}\nonumber\\
&(g)  \hspace{1cm} s_{u}[k, n]  \in  \{0, 1\}, \,\, t_{uv}[k, n]  \in  \{0, 1\},
\end{align}
$\forall u \in \mathcal{V}, vu \in \mathcal{E}, k\in \mathcal{K}, n\in [n'-T+1,n']$.

%\begin{eqnarray}
%\begin{array}%[llll]
%(a) & q_{u}[k, n] & \le & s_u[k, n]+\sum_{v\in {\cal V}}\sum_{j=0}^{D_{u}[k]}\,t_{vu}[k, n+j]\nonumber\\
%(b) & s_u[k, n]    & \le & s_u[n-1]+\sum_{v\in {\cal N}_u}t_{vu}[k, n-1]\nonumber\\
%\end{array}
%\end{eqnarray}
The above constraints reflect the storage and flow constraints \cite{llorca2013dynamic}:\\
(a) ensures that if object $k$ is requested by node $u$ at time slot $n$,
then $k$ either is in the cache of node $u$ at time $n$ or needs to
be received by node $u$ from a neighbor node $v \in {\cal N}_u$ within $D_u[k]$ time slots;\\
(b) assures that if $k$ is being cached at node
$u$ at time $n$, then $k$ either was in the cache of $u$ at time $n-1$
or was received by node $u$ from a neighbor node  $v \in {\cal N}_u$
at time $n-1$;\\
(c) assures that if object $k$ is received by
node $u$ from a neighbor node $v \in {\cal N}_u$ at time $n$, then $k$ either
was in the cache of $v$ at time $n-1$ or was received by node $v$
from a neighbor node $w \in {\cal N}_v$ at time $n-1$;\\
(d) describes the initial condition constraints that assure that each
node u always stores the objects that it hosts as a repository,
${\cal K}_u[n]$, and at $n=0$ nothing else;\\
(e) and (f) define the storage and transport capacity constraints;\\
(g) states the binary nature of the network configuration (storage and transport) variables.\\

\begin{figure}[t]
\centering
\includegraphics[width=8.5cm]{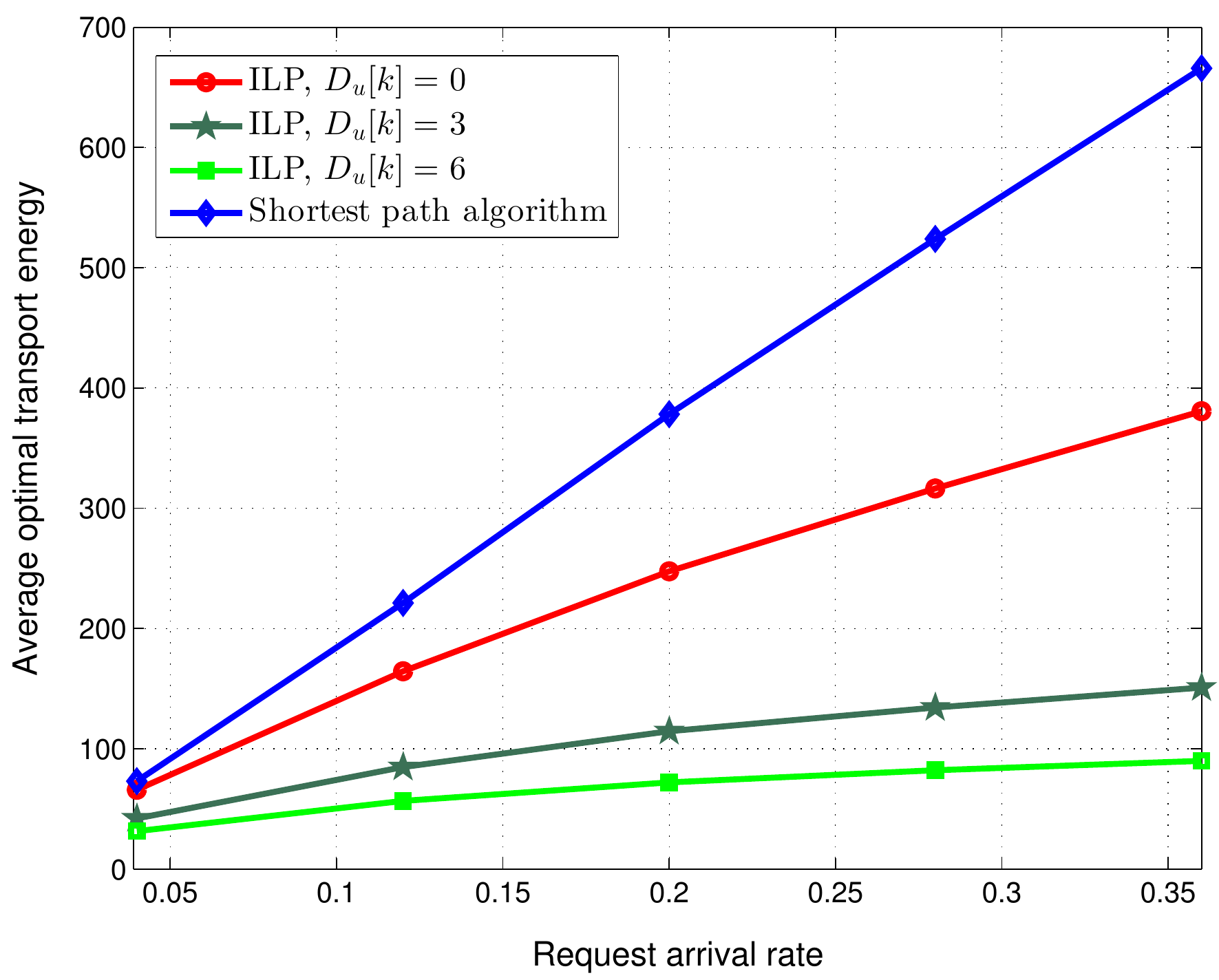}
\caption{Average optimal transport energy vs. the request arrival rate.}\label{fig:caching}
\end{figure}

To simplify the solution of the above problem, we let the entries of vector $\mathbf{x} _{T}$ to be real variables in $[0, 1]$. A numerical example resulting from our relaxed formulation is shown in Fig. \ref{fig:caching} where we
illustrate the optimal transport energy vs. the arrival request
rate. We consider a network composed of $|\mathcal{V}|=10$ nodes and $|\mathcal{K}|=4$ information objects to be  transported,
by setting $T=25$, $\vartriangle\tau=1$s, $T_{uv}=2$Mb, and $S_u=4$. We considered, for simplicity, no knowledge of popularity and same transportation costs over all links.
To better evaluate the effect of the transport energy,  we neglected the storage energy $E_{st}$ term in the integer linear program (ILP)  (\ref{eq:caching_prob}), by assuming that only three repository nodes store the information objects for all time.
As a benchmark method, we consider the shortest path algorithm, which at each request forwards the desired content along the shortest path.
It can be noted that the relaxed ILP method  yields a considerable performance gain with respect to the shortest path algorithm: moreover, the improvement  grows as the maximum delivery time $D_{u}[k]$ (set equal for each $k$)
increases, due to the greater degrees of freedom
of the algorithm.

%To meet stringent latency constraints, as required in some of 5G services, caching must be {\it proactive}, i.e. based on data analytics extracted from the data \cite{bacstuug2015big}.

\section{Graph-based resource allocation}
\label{Graph-based learning}
Enabling proactive resource allocation strategies is a key feature of 5G networks. Proactivity is rooted on the capability to predict users' behavior. Proactive caching is one example where the prediction is based on learning the popularity matrix. But of course caching is not the only network aspect that can benefit from learning. Radio coverage is one more case where learning maps of the radio environment may be useful to ensure seamless connectivity to moving users, possibly keeping the smallest number of access points active to save energy. This requires prediction of users' mobility and the capability to build Radio Environment Maps (REM) \cite{Giannakis011}. Building a REM is also a key step to enable cognitive radio \cite{Giannakis011}, \cite{yilmaz2013radio}, \cite{Giannakis017}. Balancing data traffic across the network is another problem that could take advantage of the capability to predict data flows exploiting spatio-temporal correlation (low-rank) \cite{mardani2013robust}, \cite{xu2015mining}.

\subsection{Radio environment map}
In this section, we show how graph-based representations can be useful to build a REM from sporadic measurements. Graph-based representations play a key role in many machine learning techniques, as a way to formally take into account all similarities among the entities of an interconnected system. In the signal processing community, there is a growing interest in methods for processing signals defined over a graph, or graph signal processing (GSP), for short \cite{shuman2013emerging}.
%\subsection{Radio environment map}
We show now an application of GSP to recovering the REM in a urban environment from sporadic measurements collected by mobile devices.
The goal is to reconstruct the field over an ideal grid, built according to the city map, starting from observations taken over a subset of nodes. We use a graph-based approach to identify patterns useful for the ensuing reconstruction from sparse observations. More specifically, given a set of $N$ points in space, whose coordinate vectors are $\mathbf{r}_i$ and denoting with $E_i$ the field measured at node $i$, we define the coefficients of the adjacency matrix $\mathbf{A}$ as $a_{ii}=0$ and
\begin{equation}
a_{i,j}=
\begin{cases}
&e^{-\frac{|E_i-E_j|^2}{2\sigma^2}}, \quad {\rm if}\, ||\mathbf{r}_i-\mathbf{r}_j||^2\leq R_0\\ \nonumber
&0, \quad \quad \quad \quad \quad {\rm otherwise}\\
\end{cases}, i \neq j,
\end{equation}
where $\sigma$ and $R_0$ are two parameters used to assess the similarity of two nodes: $\sigma$  is a variable used to establish the interval of values in the e.m. field within which two nodes are assumed to sense a {\it similar} value; $R_0$ is the distance within which two nodes are assumed to be neighbors. Building matrix $\mathbf{A}$ requires some prior information on the field that can be either acquired through time from measurements or it may be inferred from ray-tracing tools.
%The graph structure is built along city streets connecting pair of points according to their distances and weighting their connections according to the signal difference. Thus, it is clear the necessity of prior information about the signal, the e.m. field in this case, obtained with the help of ray-trancing tools or collecting real data thanks to measurements campaigns. With these information it is possible represent the graph in a mathematical way using the similarity/adjacency matrix $\mathbf{A}$. Denoting by $E_i$ the field at the $i$-th node, $\sigma^2$ the variance of the field, and $r_i=[x_i; y_i]$ the coordinate of the $i$-th node, the entry of adjacency matrix is
%The adjacency matrix captures the similarity between points for each active Base Station (BS), and from this matrix
From the adjacency matrix $\mathbf{A}$, we build the Laplacian matrix
\begin{equation}
\mathbf{L}=\mathbf{D}-\mathbf{A}
\end{equation}
where $\mathbf{D}$ is the diagonal matrix whose $i$-th entry is the degree of node $i$: $d_{i}=\sum_{j=1}^N a_{ij}$.
Taking the eigendecomposition of $\mathbf{L}$
\begin{equation}
\mathbf{L}=\mathbf{U}\mathbf{\Lambda}\mathbf{U}^T
\end{equation}
we have a way to identify the principal components of the field. It is well known from spectral graph theory \cite{von2007tutorial}, in fact, that  the eigenvectors associated to the smallest eigenvalues of $\mathbf{L}$ identify clusters, i.e., well connected components. Hence, the eigenvectors associated to the smallest eigenvalues of the Laplacian matrix built according to the above method are useful to identify patterns in the e.m. field. Denoting with $\mathbf{u}_k$ the eigenvector associated to the $k$-th eigenvalue, the useful signal $\mathbf{x}$ can then be modeled as the superposition of the $K$ principal eigenvectors:
\begin{equation}
\label{BL}
\mathbf{x}=\sum_{k=1}^K\mathbf{u}_k\, s_k:=\mathbf{U}_K\,\mathbf{s},
\end{equation}
with $K<N$ to be determined from measurements and $\mathbf{U}_{K}:=[\mathbf{u}_1,\ldots,\mathbf{u}_K]$.

In the GSP literature, a signal as in (\ref{BL}), with $K<N$, is called a {\it band-limited} signal over the graph. In general, a real signal is never perfectly bandlimited, but it can be approximately bandlimited. Having a band-limited model is instrumental to establish the condition for the recovery of the entire signal from a subset of samples \cite{tsitsvero2015signals}.

In a real situation, it is typical to have several access points whose radio coverage areas overlap. For each access point, we can build a dictionary using the method described above, using for the e.m. field a ray-tracing algorithm.  We denote by $\mathbf{U}_K^{(m)}$ the dictionary built when only AP $m$ is active.  At any given time frame, only a few AP's are active. Therefore, the overall map can be written as
\begin{equation}
\mathbf{x}=\sum_{m=1}^M\sum_{k=1}^K\mathbf{u}_k s_k^{(m)}:=\sum_{m=1}^M\,\mathbf{U}_K^{(m)}\mathbf{s}^{(m)}:=\mathbf{U}\mathbf{s},
\end{equation}
where $M$ is the number of AP's covering the area of interest (not all of them necessarily active at the same time), $\mathbf{U}:=(\mathbf{U}_K^{(1)}, \ldots, \mathbf{U}_K^{(M)})$ and  $\mathbf{s}:=(\mathbf{s}^{(1)}; \ldots; \mathbf{s}^{(M)})$ is sparse. The observed signal typically consists in a limited number of measurements collected along the grid. We may write the observed signal as:
\begin{equation}
\mathbf{y}= \mathbf{\Sigma}\,\sum_{m=1}^M\,\mathbf{U}_K^{(m)}\mathbf{s}^{(m)}=\mathbf{\Sigma}\mathbf{U}\mathbf{s},
\end{equation}
where $\mathbf{\Sigma}$ is a diagonal selection matrix, whose $i$-th entry is one if node $i$ is observed, and zero otherwise. The recovery of the overall radio coverage map can then be formulated as a sparse recovery problem. We used Basis Pursuit (BP), which implies solving the following convex problem:\\
\begin{align}
\hat{\mathbf{s}}&={\rm arg}\min_{\mathbf{s}} \|\mathbf{s}\|_1\nonumber\\
& {s.t.}\,\,\mathbf{y}=\mathbf{\Sigma}\mathbf{U}\mathbf{s}
\end{align}
and then we used $\hat{\mathbf{x}}=\mathbf{U}\,\hat{\mathbf{s}}$.\\

\begin{figure}[h]
\centering
\includegraphics[width=\textwidth]{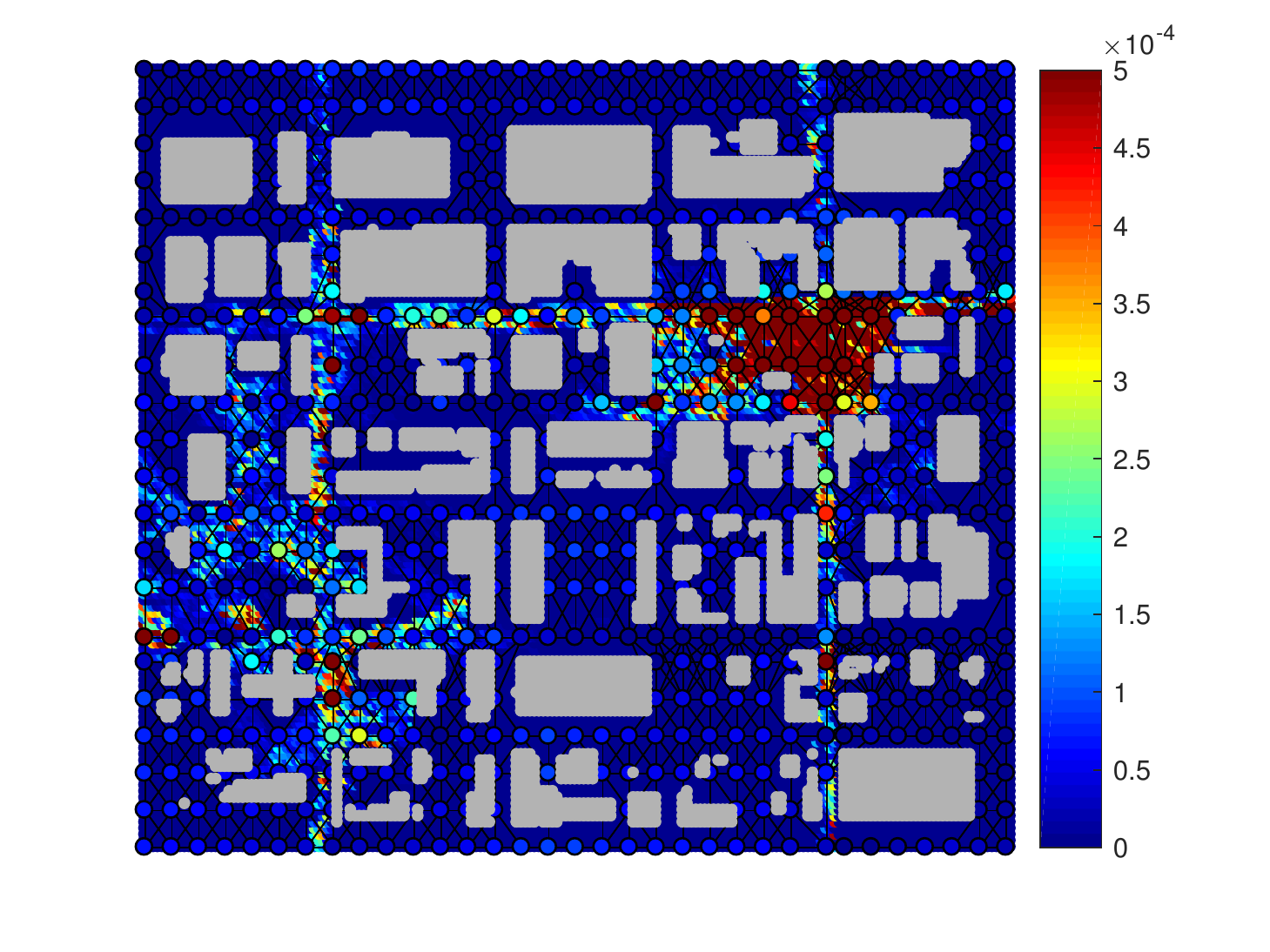}
\caption{Example of reconstructed e.m. field.}
\label{fig:cartography}\end{figure}

An example of reconstruction using BP is shown in Fig.\ref{fig:cartography}.
The grid is composed  of $N=547$ nodes and the number $M$ of AP's covering the city area illustrated in the figure is $4$. The AP's are located in the south-east, north-east, north-west and south-west side of the examined area. The number of measurements is $115$. Measurement noise is considered negligible. We assumed a bandwidth $K=40$, equal for all AP's. The background (continuous) color is the map ground-truth, obtained using the ray-tracing tool Remcom Wireless InSite 2.6.3 \cite{remcomRay}. The colors on each vertex of the grid represent the reconstructed value. Comparing each node color with the background, we can testify the goodness of the method to reconstruct the overall map.  The Normalized Mean Square Error (NMSE), measured as the square norm of the error, normalized by the square norm of the true signal, in this example, is  $NMSE=0.018$. The quality of the reconstruction depends on the number of measurements and on the assumption on the bandwidth. Clearly, the larger is the bandwidth, the better is the reconstruction, but the larger is also the number of measurements to be taken to enable the reconstruction. This suggests that the choice of the bandwidth must come from a trade-off between accuracy and complexity.
% \section{Graph-based resource allocation}
% \label{Graph-based resource allocation}
% In Section \ref{Joint optimization of communication and computation} we showed that a joint allocation of computation and communication resources can provide better performance than a disjoint one in computation offloading applications. In that section, we assumed a predefined assignment of users to access points and MEC servers. In this section we address the assignment problem and we propose a method based on matching theory.
\subsection{Matching users to $3C$ resources}
\label{Graph-based resource allocation}
In Section \ref{Joint optimization of communication and computation} we motivated the use of a joint allocation of computation and communication resources in computation offloading. We also incorporated the assignment rule between UE, AP, and MEC within the overall optimization problem. The resulting formulation yields better performance than a disjoint formulation, however it is also computationally demanding because it involves the solution of a mixed-integer programming problem.

A possible way to overcome this difficulty is to simplify the rule for associating UE's to AP and MEC. One possibility is to resort to {\it matching theory}, a low complexity tool used to solve the combinatorial problem of matching players from different sets, based on their preferences. Matching theory can be seen as the problem of finding a bipartite graph connecting two sets, depending on the preference lists. Matching theory has already been proposed in \cite{Gu2015} for resource allocation in multi-tiered wireless heterogeneous architectures, with applications to cognitive radio networks, heterogeneous small-cell-based networks and Device-to-Device communications (D2D). In \cite{Li2017}, a multi-stage matching game is used in the C-RAN context to assign Radio Remote Heads (RRH), Base Band Units (BBU) and computing resources for computation offloading, aimed at minimizing the refusal ratio, i.e. the proportion of offloading tasks that are not able to meet their deadlines. A well-known matching problem is the college admission game presented in \cite{Gale1962}, where a Deferred-Acceptance (DA) algorithm is proved to converge to a stable matching with extremely low complexity. The key initial step of matching theory is to establish a preference rule. For instance, in \cite{Saad2014} the users' preferences are defined as the $R$-factor, which captures both Packet Success Rate (PSR) and wireless delay. However, as pointed out in \cite{Saad2014}, the complexity of this algorithm increases considerably when dealing with interdependent preferences, i.e. when the preference of a user is affected by the acceptance of the others. This is indeed the case of user association in wireless networks, because, continuing in the example defined above, the $R$ factor of a user changes as other users get accepted by the same AP. To overcome this problem, the authors of \cite{Saad2014} divide the game into two interdependent subgames:
\begin{enumerate}
\item An admission matching game with $R$-factor guarantees, depending on the maximum delay experienced at each access point;
\item A coalitional game among access points, where the coalitions are sets of AP's and associated users.
\end{enumerate}
In particular, a user assigned to a certain AP $a$ through the first subgame, could prefer to be matched to another AP $b$, since the utility functions change as users get admitted. Then, a user $k$ requests to be transferred from $a$ to $b$ if it improves its $R$-factor. The transfer is accepted if and only if:
\begin{enumerate}
\item  The access point $b$ does not exceed its quota (maximum number of admitted users);
\item The social welfare (sum of the $R$-factors of the two coalitions) is increased.
\end{enumerate}
Starting from an initial partition (sets of coalitions) obtained with the deferred acceptance algorithm, the algorithm in \cite{Saad2014} converge to a final partition that is also Nash-stable. In the holistic view of $3C$ resources, other utility functions can be used to take into account all the three aspects of $3C$: communication, computation, and caching. For instance, additional parameters to be taken into account are the computational load on MEC servers in case of computation offloading and the amount of storage for caching.\\

One more example where graph theory can be used is load balancing. In fact, especially in view of the dense deployment of access points, there is a high probability that the load, either data rate, computational load or storage, can be highly unbalanced throughout the network \cite{vu2017joint}. One possibility to balance the situation is to split the networks in many non-overlapping clusters. A cluster head is then elected in each cluster and it enforces a balance within the cluster. Then, balancing across clusters is achieved by repeated clustering and balancing steps. A possible way to do clustering is to use spectral clustering, which starts from the creation of a similarity (adjacency) matrix. In this case, as suggested in \cite{samarakoon2014dynamic}, it could be useful to include in the construction of the adjacency matrix a {\it dissimilarity} measure that assesses how much two nodes are unbalanced. In this way, the ensuing clustering tends to put together nodes that are close but unbalanced so that the resulting in-cluster balancing will be more effective.

\section{Network reliability}
\label{Network reliability}
The edge-cloud architecture described in Section \ref{Holistic view of communication, computation and caching} clearly builds on the reliability of the network connectivity. However, in practice, the presence of a link between a pair of nodes is subject to random changes. In a wireless communication system, for instance, it is typical to have random link failures due to fading. With mmWave communications, link failures are typically even more pronounced because of blocking due to obstacles between transmit and receive devices.
%Similarly, in a point cloud, the association of an edge to a pair of points can also be a random event, because of  imperfect information in the rule used to decide whether to establish an edge or not.
The goal of this section is to build on graph-based representations to assess the effect of random failure on a limited number of edge on macroscopic network parameters, such as, for example, connectivity. We build our study  on a small perturbation analysis of the eigendecomposition of the  Laplacian matrix describing the graph, as suggested in \cite{Ceci2018}. An outcome of our analysis is the identification of the most critical links, i.e. those links whose failure has a major effect on some network macroscopic features, such as connectivity.

A small perturbation analysis of the eigen-decomposition of a matrix is a classical problem that has been studied since a long time, see, e.g. \cite{wilkinson1988}, \cite{stewart1973}.
In this section we focus on the small perturbation analysis of the eigendecomposition of a perturbed Laplacian $\bL+\delta\bL$, incorporating an original graph Laplacian $\bL$ plus the addition or deletion of a small percentage of edges. We consider a graph composed of $N$ vertices, so that the dimension of $\bL$ is $N \times N$. We denote by $\tilde{\lambda_i}=\lambda_i+\Delta\lambda_i$ the perturbed $i$-th eigenvalue and  by $\tilde{\bu_i}=\bu_i+\Delta\bu_i$ the associated perturbed eigenvector.
If only one link fails, let us say link $m$, the perturbation matrix can be written as $\delta \bL(m)=-\ba_{m}\ba_{m}^T$, where $\ba_m=[a_{m_{1}} \cdots a_{m_{n}}]^T$ is a column vector of size $N$ that has all entries equal to zero, except the two elements $a_{m}(i_m)=1$ and $a_{m}(f_m)=-1$, where $i_m$ and $f_m$ are the initial and final vertices of the failing edge $m$. In case of addition of a new edge, the perturbation matrix is simply the opposite of the previous expression, i.e. $\delta \bL(m)=\ba_{m}\ba_{m}^T$. It is straightforward to see that the perturbation of the Laplacian matrix due to the simultaneous deletion of a small set of edges is simply $\delta \bL= -\sum_{m \in {\cal E}_p} \ba_m \ba_m^T$ where ${\cal E}_p$ denotes the set of perturbed edges.
 %We are concerned with undirected graphs, so that we consider the simple case of of real symmetric matrices.
The perturbed eigenvalues and eigenvectors $\tilde{\lambda_i}$ and $\tilde{\bu_i}$, in the case where all eigenvalues are distinct and the perturbation affects a few percentage of links,
are related to the unperturbed values  $\lambda_{i}$ and $\bu_{i}$ by the following formulas \cite{wilkinson1988}:
 %\footnote{In this paper, we assume that all eigenvalues have multiplicity one.}
\begin{equation}
\label{lambdaP}
\tilde{\lambda_i}\simeq\lambda_{i}+\bu_{i}^{T}\delta \bL\, \bu_{i}
\end{equation}
\begin{equation}
\label{icsP}
\tilde{\bu_i}\simeq\bu_{i}+ \sum_{j\neq i}\frac{\bu_{j}^T\delta \bL\, \bu_{i}}{\lambda_{i}-\lambda_{j}}\bu_{j}.
\end{equation}
In particular, the perturbations due to the failure of a generic link $m$ on the $i$-th eigenvalue and associated eigenvector are:
\begin{align}
\label{Delta_lambda}
\Delta\lambda_i(m)&=\bu_{i}^{T}\delta \bL(m) \bu_{i}=-\bu_{i}^{T}\ba_{m}\ba_{m}^T\bu_{i}= \nonumber\\
& = -||\ba_{m}^T\bu_{i}||^2=-[u_{i}(f_m)-u_{i}(i_m)]^2
\end{align}
and
\begin{align}
\label{deltaU}
\Delta \bu_i(m)&=\sum_{j\neq i} \frac{\bu_j^T\delta L(m)\bu_i}{\lambda_i-\lambda_j}\bu_j
=-\sum_{j\neq i} \frac{\bu_j^T\ba_m \ba_m^T \bu_i}{\lambda_i-\lambda_j}\bu_j\nonumber\\&=\sum_{j\neq i} \frac{[u_j(i_m)-u_j(f_m)][u_i(f_m)-u_i(i_m)]}{\lambda_i-\lambda_j}\bu_j.
%\nonumber\\
%&= [u_{j}(i_m)-u_{j}(f_m)][u_{i}(f_m)-u_{i}(i_m)].
\end{align}
\textcolor{black}
Within the limits of validity of first order perturbation analysis, the overall perturbation resulting from the deletion of multiple edges is the sum of all the perturbations occurring on single edges:
\begin{equation}
\Delta\lambda_i=\sum_{m\in{\cal{E}}_p}\Delta\lambda_i(m),
\end{equation}
where ${\cal{E}}_p$ denotes the set of perturbed edges.
In their simplicity, the above formulas  capture some of the most relevant aspects of perturbation and their relation to graph topology. In fact, it is known from spectral graph theory, see e.g., \cite{von2007tutorial}, that the entries of the Laplacian eigenvectors associated to the smallest eigenvalues tend to be smooth and assume the same sign over vertices within a cluster, while they can vary arbitrarily across different clusters.  Taking into account these properties, the above perturbation formulas (\ref{lambdaP})-(\ref{deltaU}) give rise to the following interpretations:
\begin{enumerate}
\item the edges whose deletion causes the largest perturbation are inter-cluster edges;
\item given a connected graph, the eigenvector associated to the null eigenvalue does not induce any perturbation on any other eigenvalue/eigenvector, because it is constant;
\item the eigenvector perturbation is larger for quantities (either eigenvalues or eigenvectors) associated to eigenvalues very similar to each other (recall that formulas (\ref{lambdaP}) and (\ref{icsP}) hold true only for distinct eigenvalues).
\end{enumerate}

\subsection{A new measure of edge centrality}
Based on the above derivations, we propose a new measure of edge centrality, which we call {\it perturbation centrality}. We assume a connected undirected graph. If we denote by $K$ the number of clusters in the graph and by $\Delta \lambda_i(m)$ the perturbation of the $i$-th eigenvalue due to the deletion of edge $m$, we define the topology perturbation centrality of edge $m$  as follows \cite{Ceci2018}:
\begin{equation}
\mathsf{p}_K(m):=\sum_{i=2}^K\, |\Delta \lambda_i(m)|.
\end{equation}
The summation starts from $i=2$ simply because, from (\ref{lambdaP}), the perturbation induced by the deletion of any edge on the smallest eigenvalue is null. The above parameter $\mathsf{p}_K(m)$ assigns to each edge the perturbation that its deletion causes to the overall network connectivity, measured
as the sum of the  $K$ smallest eigenvalues of the Laplacian matrix \cite{von2007tutorial}. This parameter is particularly relevant in case of modular graphs, i.e. graphs evidencing the presence of clusters. In such a case, it is well known from spectral clustering theory \cite{von2007tutorial} that the smallest eigenvalues of the Laplacian carry information about the number of clusters in a graph.

\begin{figure}[h]
\centering
\includegraphics[width=0.85\textwidth]{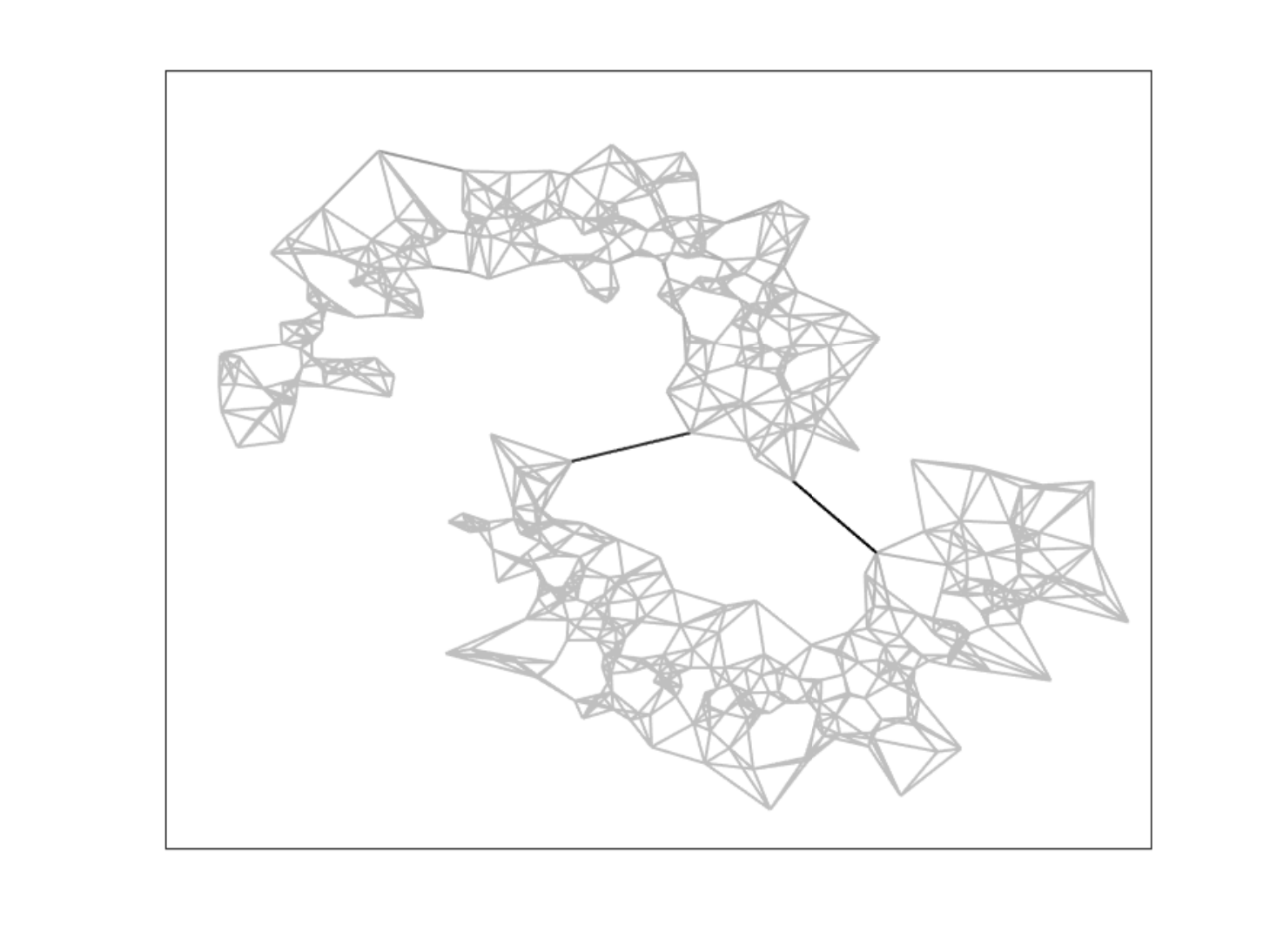}
\caption{Example of perturbation centrality measure.}
\label{fig:centrality}
\end{figure}

In Fig. \ref{fig:centrality} we report an example of modular graph, obtained by connecting two clusters through a few edges. The {\it{perturbation centrality} }is encoded in the color intensity of each edge. It is interesting to see that the edges with the darkest color are, as expected, the ones connecting the two clusters.

\subsection{Application: Robust information transmission over wireless networks}
\label{applications1}
\begin{figure}[h]
\centering
\includegraphics[width=0.8\textwidth]{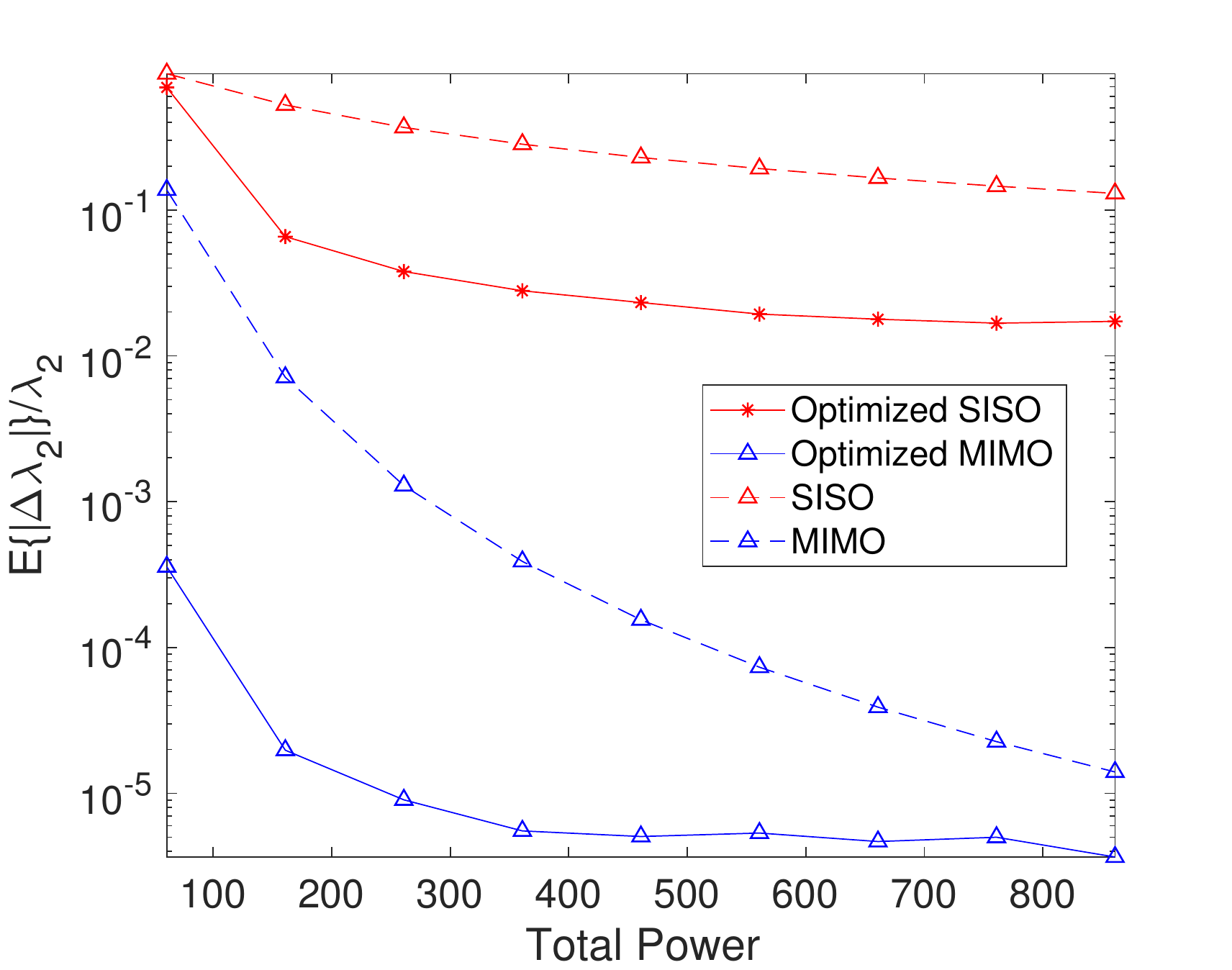}
\caption{Expected perturbation of algebraic connectivity vs. total power.}
%\caption{provA CAPTION}
\label{relaPet}
\end{figure}

Now we apply our statistical analysis to optimize the resource (power) allocation over a wireless network in order to make the network robust against random link failures. We consider a wireless communication network with $M$ links, where each link is subject to a random failure because of fading or blocking. Every edge is characterized by an outage probability $P_{out}(m), m=1,\ldots,M$. We suppose the failure events over different links to be independent of each other. We consider first a single-input-single-output (SISO) Rayleigh flat fading channel for each link. In such a case,  the channel coefficient $h$ is a complex Gaussian random variable (r.v.) with zero mean and circularly symmetric. Hence, the r.v. $\alpha=|h|^2$ has an exponential distribution. Denoting with $F_n(x; \lambda)$ the cumulative distribution function (CDF) of a gamma random variable $x$ of order $n$, with parameter $\lambda$, the CDF of $\alpha$ can then be written as $F_1(\alpha; \lambda)$. We also denote with $C=\log_2(1+|h|^2\rho)$ the link capacity (in bits/sec/Hz), where $\rho=\frac{P_T(m)}{\sigma_n^2 r_m^2}$ is the signal-to-noise ratio (SNR), $P_T(m)$ is the transmitted power over the $m$-th link, $\sigma_n^2$ is the noise variance, and $r_m$ the distance covered by link $m$. Denoting by $R$ the data rate, the outage probability $P_{out}(m)$ is defined as:
\begin{align}
\label{Pout(m)}
P_{out}(m)&=Pr\{C<R\}=Pr\{\log_2(1+|h|^2\rho)<R\}\\ \nonumber
&= Pr\{|h|^2<\frac{2^R-1}{\rho}\}\\ \nonumber
&=\int_0^{\frac{2^R-1}{\rho}}\lambda e^{-\lambda\alpha}d\alpha=F_1 \left(\frac{2^R-1}{\rho}; \lambda\right)=1-e^{-\frac{\lambda}{\rho}(2^R-1)}.
\end{align}
Since the CDF of $\alpha$ is invertible, it is useful to introduce its inverse. In particular, if $y=F_n(x; \lambda)$, we denote its inverse as $x=F_n^{-1}(y; \lambda)$. Expression (\ref{Pout(m)}) can then be inverted to derive the transmit power $P_T(m)$ as a function of the outage probability:
\begin{equation}
\label{funDiP}
P_T(m)=-\frac{\lambda\sigma_n^2 r_m^2(2^R-1)}{\log(1-P_{out}(m))}=\frac{\sigma_n^2 r_m^2(2^R-1)}{F_1^{-1}(P_{out}(m); \lambda)}.
\end{equation}
The small perturbation statistical analysis derived above can be used to formulate a robust network optimization problem. We assess the network robustness, in terms of connectivity,  as the ability of the network to give rise to small changes of connectivity, as a consequence of a small number of edge failures. The network connectivity is measured by the second smallest eigenvalue of the Laplacian, also known as the graph {\it algebraic connectivity}. This parameter is known to provide a bound for the graph conductance \cite{newmanBook}. Our goal now is to evaluate the transmit powers $P_T(m)$, or equivalently, through (\ref{funDiP}), the outage probabilities, that minimize the average perturbation of the algebraic connectivity, subject to a cost function on the total transmit power $P_{T_{max}}$ of the overall network. In formulas, we wish to solve the following optimization problem:
\begin{equation}
\label{optP}
\begin{array}{lll}
&\underset{\mathbf{P}_{out}} \min \quad  \displaystyle \sum_{m\in {\cal{E}}}  {\mathbb{E}\{|\Delta\lambda_2(m)|\}}\medskip\\
&s.t.  \quad \underset{m\in\mathcal{E}}{\sum}P_{T}(m)\leq P_{T_{max}}\medskip\\ \nonumber
& \quad \quad P_{out}(m)\in [0,1],\,\forall\, m \in \mathcal{E}.
\end{array}
\end{equation}
Using equation
(\ref{Delta_lambda}) and (\ref{funDiP}),
we can rewrite the optimization problem explicitly in terms of the outage probabilities $P_{out}(m)$ as:
%From equation (\ref{lambdaP}), we have a simple formula to assess the eigenvalue perturbation. Therefore, the optimal transmit powers, or edges failure probabilities, can be found as the result of an optimization problem, where it is minimized the weighted sum of the perturbation produced by each link failure, each one weighted by the corresponding probability, under the constraint on the overall cost function. The formulation becomes:
\begin{equation}
\begin{array}{ll}
& \underset{\mathbf{P}_{out}}{\min} \quad \displaystyle \sum_{m\in\mathcal{E}} P_{out}(m){[u_{2}(i_m)-u_{2}(f_m)]^2}\\ \nonumber
&  s.t.\quad \quad \quad \quad \quad \quad \quad  \quad \quad \quad \quad \quad \quad \quad \quad \quad \quad \quad \quad (\mathcal{Q})\nonumber \\
& \quad \quad \quad \,\underset{m\in\mathcal{E}}{\sum} \frac{r_m^2}{F_1^{-1}(P_{out}(m); \lambda)}\leq C_{max} \medskip \\\nonumber
& \quad \quad \quad\,P_{out}(m)\in [0,1],\,\forall\, m \in \mathcal{E}\nonumber
\end{array}
\end{equation}
where $C_{max}:=\frac{P_{T_{max}}}{\sigma_n^2 (2^R-1)}$.\\
Problem ($\mathcal{Q}$) is non-convex because the constraint set is not convex. However, if we perform the change of variable $t_m:={1}/{F_1^{-1}(P_{out}(m); \lambda)}=-\lambda/\log(1-P_{out}(m)), m=1,\ldots,M$, the first constraint becomes linear. The objective function becomes non-convex. However, if we limit the variability of the unknown variables to the set $t_m \geq \lambda/2, \, \forall\, m$, the objective function becomes convex, so that the original problem
converts into the following convex problem:
\begin{equation}
\begin{array}{lll}
&\underset{\mathbf{t}}{\min}\quad &\underset{m\in \mathcal{E}}{\sum}F_1(\frac{1}{t_m}; \lambda) {|\Delta\lambda_2(m)|}= \underset{m\in \mathcal{E}}{\sum}(1-e^{-\frac{\lambda}{t_m}}){|\Delta\lambda_2(m)|}\medskip\\
&s.t. \quad &\underset{m\in \mathcal{E}}{\sum} r_m^2 t_m\leq C_{max} \quad \quad \quad \quad \quad \quad \quad \quad \quad \quad \quad (\mathcal{Q}_1) \medskip\\
&\quad \quad & t_m \geq \frac{\lambda}{2},\quad \forall m\in \mathcal{E}.
\end{array}
\end{equation}
% \textcolor{red}{In the following paragraph we analyze the convex problem in SISO and MIMO scenario, how change the bounded region and the comparison of the performance. Then we will suggest a intermediate approach which present SISO communication system for that links which are in a lower critical status, and MIMO system for the worst performance links in the network connectivity point of view.}
We can now generalize the previous formulation to the Multi-Input Multi-Output (MIMO) case, assuming multiple independent Rayleigh fading channels. One fundamental property of MIMO systems is the diversity gain, which makes them more robust against fading with respect to SISO systems \cite{barbarossa2003}. In fact, different performance can be obtained depending on the number of antennas on the transmitting sides $n_T$ and receiving sides $n_R$ exploiting the diversity gain. In a MIMO system whit $n=n_T\times n_R$ statistically independent channels, denoting by $h_{ij}$ the coefficient between the $i$-th transmit and the $j$-th receive antenna, the pdf of the random variable $\alpha:=\sum_{i=1}^{n_T}\sum_{j=1}^{n_R}|h_{ij}|^2$ is the Gamma distribution:
\begin{equation}
P_{A}(\alpha)=\frac{\lambda^n}{(n-1)!}\,\alpha^{n-1}e^{-\lambda \alpha}
\end{equation}
and we denote by $F_n(\alpha; \lambda)$ its cumulative distribution function (CDF), with parameters $n$ and $\lambda$.
Proceeding similarly to the SISO case, the optimization problem can be formulated as
%outage probability is now computed taking to account the Gamma pdf as follows:
%\begin{equation}
%\begin{array}{lll}
%&P_{out}(m)=Pr\{C<R\}=\Pr\{h^2<\frac{2^R-1}{\rho}\}=\\ \nonumber
%&\int_0^{\frac{2^R-1}{\rho}}\frac{1}{b^n(n-1)!}\alpha^{n-1}e^{-\frac{\alpha}{b}}d t :=F(\frac{(2^R-1)\sigma_n^2r^2}{P_T}|n,b)
%\end{array}
%\end{equation}
%where $F(x|a,b)$ denotes the CDF of $\alpha$. As in the previous case the $P_{out}(m)$ was equal to CDF of exponential distribution, now it is equal to the Gamma cumulative distribution function which has not a closed form. Nonetheless it is still possible to express the transmitted power as a function of $P_{out}$ using $F^{-1}(P_{out}(m)|a,b)$ that is the inverse of the CDF:
%\begin{equation}
%f(P_{out}(m))=P_T=\frac{\sigma_n^2 r^2(2^R-1)}{F^{-1}(P_{out}(m)|n,b)}
%\end{equation}
%Now we can formulate an equivalent convex problem to SISO case, where the bounded region here is dependent on the degrees of freedom $a$ of the Gamma distribution. In fact, to linearize the constraint, that in MIMO system is become $\underset{k\in\mathcal{E}}{\sum} \frac{\sigma_n^2r^2(2^R-1)}{F^{-1}(P_{out}(m)|n,b)}\leq C_{max}$, we set $t_m=\frac{1}{F^{-1}(P_{out}(m)|n,b)}$ and we found a bounded region $\bt \geq 1/b(n-1)$ inside with the objective function where we operate the variable substitution is convex. The problem formulation becomes the following:
\begin{equation}
\begin{array}{lll}
&\underset{\mathbf{t}}{\min}\quad &\underset{m\in \mathcal{E}}{\sum}F_n(\frac{1}{t_m}; \lambda) {|\Delta\lambda_2(m)|}\\
&s.t. \quad &\underset{m\in \mathcal{E}}{\sum} r_m^2 t_m\leq C_{max}	 \quad \quad \quad \quad \quad \quad (\mathcal{Q}_2)\\
&\quad \quad & t_m\geq \lambda/(n+1),\quad \forall m\in \mathcal{E}
\end{array}
\end{equation}
where the constraint on the variables $t_m$ has been introduced to make the problem convex.
%\begin{figure}
%\includegraphics[width=0.5\textwidth]{confrotoMIMOSISO.eps}
%\caption{Comparisons between SISO (red curves) and MIMO ($n=4$) systems (blue curves), with and without optimization.}
%\label{relaPet}
%\end{figure}
Indeed, problem ${\cal{Q}}_1$ is a special case of problem ${\cal{Q}}_2$, when $n=1$. An interesting result about the convexity of problem ${\cal{Q}}_2$ is that the bounding region increases with the number of independent channels.\\
As a numerical example, we considered a connected network composed by two clusters, with a total of $|{\cal{E}}|=761$ edges and four bridge edges between the two clusters. For the sake of simplicity, we assumed the same distances $r_m$ over all links.
In Fig. \ref{relaPet}, we compare the expected perturbations of the algebraic connectivity, normalized to the nominal value $\lambda_2$, obtained using our optimization procedure or using the same power over all links, assuming the same overall power consumption. We report the result for both SISO and MIMO cases. From Fig. \ref{relaPet}, we can observe a significant gain in terms of the total power necessary to achieve the same expected perturbation of the network algebraic connectivity. We can also see the advantage of using MIMO communications, at least in the case of statistically independent links.

\section{Conclusions}
\label{Conclusions}
In this chapter we have described some of the aspects of the edge-cloud architecture, a framework proposed to bring cloud and communication resources as close as possible to mobile users to reduce latency and achieve a more efficient usage of the available energy. From the edge-cloud perspective, we have motivated a holistic view that aims at optimizing the allocation of communication, computation and caching resources jointly. Within this framework, graph-based representations play a key role. In this chapter, we considered just a few cases where these representations can provide a valid and innovative tool for an efficient deployment of the edge-cloud system. As it happens in most engineering problems, big potentials come with big challenges. One of these is complexity. To take full advantage of graph representations, there is the need for devising efficient distributed computational tools to analyze graph-based signals. Furthermore, we believe that graph representations are only the beginning of the story, as they are built incorporating only pairwise relations. More sophisticated tools may be envisaged by enlarging the horizon to include multi-way relations, using for example simplicial complexes or hypergraphs, as suggested in \cite{barbarossa2016introduction}, or multilayer network representations \cite{kivela2014multilayer}, \cite{boccaletti2014structure}. Furthermore, in this work, we have basically restricted our attention to time-invariant graph representations and to linear models. Clearly, a significant improvement can be expected by enlarging the view to time-varying graphs and nonlinear models \cite{shen2017kernel}, \cite{romero2016kernel}.
\section{Acknowledgments}
The research leading to these results has been jointly funded by the European Commission (EC) H2020 and the Ministry of Internal affairs and Communications (MIC) in Japan under grant agreements Nr. 723171 5G MiEdge in EC and 0159-{0149, 0150, 0151} in MIC.
\section*{References}
\bibliographystyle{Vancouver-Numbered-Style_3_}
\bibliography{refs}

\end{document}